\documentclass[preprint,authoryear,12pt]{elsarticle}
\usepackage{graphicx}
\usepackage{amssymb}

\journal{Journal of the Mechanics and Physics of Solids}

\begin{document}

\begin{frontmatter}

\title{Creep rupture of materials: insights from a fiber bundle model with relaxation}

\author{E. A. Jagla}

\address{Centro At\'omico Bariloche, Comisi\'on Nacional de Energ\'{\i}a At\'omica, 
(8400) Bariloche, Argentina}

\begin{abstract}

I adapted a model recently introduced in the context of seismic phenomena, to study creep rupture of materials. It consists of
linear elastic fibers that interact in an equal load sharing scheme, complemented with a local viscoelastic relaxation mechanism.
The model correctly describes the three stages of the creep process, namely an initial Andrade regime of creep relaxation, an intermediate regime of rather constant creep rate, and a tertiary regime of accelerated creep towards final failure of the sample. In the tertiary regime creep rate follows the experimentally observed one over time-to-failure dependence. The time of minimum strain rate is systematically observed to be about 60-65 \% of the time to failure, in accordance with experimental observations.  In addition, burst size statistics of breaking events display a $-3/2$ power law for events close to the time of failure, and a steeper decay for the all-time distribution. Statistics of interevent times shows a tendency of the events to cluster temporarily. This behavior should be observable in acoustic emission experiments.
\end{abstract}

\end{frontmatter}

\section{Introduction}

Creep is the time deformation of a material under the application of a mechanical load that is below its overall mechanical strength.
In many cases, creep can lead to the failure of the material after a finite time after the application of the load, and this kind of creep rupture can be one of the mayor concerns in mechanical design.
Creep rupture follows a sequence of three temporal stages before complete failure occurs.
Immediately after the load is applied, there is a so-called primary creep regime in which the strain rate in the material decays as a function of time. This is known as the Andrade creep regime \citep{andrade}. The second stage is one of almost constant strain rate, that gives way eventually to the tertiary regime of accelerated creep that leads to complete material collapse. In the Andrade regime and the tertiary creep regime the strain rate is observed to follow power laws in time. In the Andrade regime, strain rate $\dot \varepsilon$ decreases as a negative power of the time elapsed since load application $\dot \varepsilon\sim t^{-p}$, whereas in the tertiary creep regime, strain rate increases as a negative power of the time remaining to failure, i.e. $\dot \varepsilon\sim (t_f-t)^{-p'}$, where $t_f$ is the time at which failure
is observed. Experimental values typically observed for the exponents $p$ and $p'$ are both close to 1 \citep{nechad_jmps,nechad_prl}. The time to failure $t_f$ depends strongly on the applied load $\sigma$. Actually, if $\sigma$ is below some threshold value $\sigma_c$, failure does not occur, and only the initial Andrade regime towards a finite deformation state is observed. Time to failure is expected to diverge continuously as $\sigma \to \sigma_c^+$, however, the precise experimental determination of this dependence suffers of the problem that the determination of $t_f$ is typically destructive, and different samples have to be used for different values of applied stress, introducing sample-to-sample fluctuations that complicate the determination of $t_f$.
The typical dependence of strain rate vs time displays another interesting feature: the time at which strain rate is minimum $t_m$, is typically a conserved fraction of the time to failure, independently of the level of applied stress, and details of the samples. \citet{nechad_jmps} find a ratio $t_f/t_m\simeq 1.58$ for a set of fifteen samples with different characteristics.
In addition to the average evolution of the creep process, characterized by the temporal evolution of the strain, or the strain rate, there is relevant information that can be extracted by analyzing the details of the temporal evolution. In fact, creep proceeds typically through bursts of activity in which the strain increases in a finite amount, followed by periods of quiescence. This pattern of activity, which is reminiscent for instance of the Barkhaussen noise in ferromagnets, or earthquakes in seismic processes, can be typically followed by recording the patterns of acoustic emission generated by the sample \citep{maes,ae2,ae}.

There is a variety of models that have been used to explain part of the phenomenology previously described about the creep rupture of materials. Most of them are based in the idea of ``fiber bundles", in which the material is considered as a set of fibers that fail individually when some threshold value of stretching is overpassed \citep{daniels}. The stress carried by fibers that break is distributed to the intact fibers of the system, and this generates a feedback process that can produce  the total failure of the system. 
The description of creep phenomena by the use of fiber bundle model requires the consideration of some time dependent process in the model. In some cases, it is assumed that the external load increases with time, usually this increase is thought to be quasistatic. This is the ``static" FBM considered for instance by \citet{pradhan_rmp}. In a slightly different approach introduced by \citet{kun_pre_2003}, the external load is kept constant, but the system is endowed with some kind of viscoelastic behavior that produce a time dependent response.

An additional effort to set up a model that describes the full phenomenology of the three regimes of creep behavior has been made by \citet{nechad_jmps,nechad_prl}.
The authors show how previous versions of FBMs fail to obtain realistic strain rate dependencies in the Andrade and tertiary creep regime. They construct upon the previous models by modifying the rheological properties of the fibers, in order to reproduce these regimes. The final conclusion is that an Eyring rheology is able to predict a power law in the Andrade and tertiary creep regimes. The crossover time $t_m$ found with this model is equal to $t_f/2$, and the dependence of time to failure with the excess stress $\sigma-\sigma_c$ is $t_f \sim (\sigma-\sigma_c)^{-1/2}$.

Despite the fact that the modeling presented by \citet{nechad_jmps,nechad_prl} is able to reproduce many of the observed properties of the creep behavior of the material, I would like to present here an alternative model that also explains many characteristics of the creep process. So first of all it is necessary to explain why we need another model of creep rupture. The model of \citet{nechad_jmps,nechad_prl} relies on the assumption that fibers respond through an ad hoc non-linear rheology. Although the assumed non-linear behavior of each fiber can be considered realistic, as it is compatible with macroscopic phenomenology of thermal creep, to apply this macroscopic behavior to presumably microscopic elements leaves the impression that 
we have not reached a really fundamental description of creep. In ddition, this model (as the viscoelastic model of \citet{kun_pre_2003}) produces
a temporal breaking sequence of fibers in a rigorous one-by-one process, i.e., avalanches of many breakings do not occur. This forces the authors in \citet{nechad_prl,nechad_jmps} not to consider the fitting of acoustic emission data. In this respect, the model presented by \citet{baxevanis_2007,baxevanis_2008} seems a bit more realistic, however it displays some non-physical behavior also (see below).
Another  additional minor points is that the model in \citet{nechad_prl,nechad_jmps} is not able to explain other robust features of the experiments, such as the already mentioned constant ratio $t_m/t_f\sim 0.6$. 

The second set of reasons to propose a different model for creep rupture comes from the fact that the `new' model I will present, is not actually new, nor devised specially to treat this problem, but is a model that has proved very successful in reproducing the characteristics of temporal sequences of seismic events, giving in particular a very realistic description of aftershocks \citep{jagla_pre,jagla_jgr}.
With rather  direct modifications, the modeling of the seismic phenomena can be adapted to model creep rupture with reasonably good results. As we will see, except for the possibility of breaking of elements, the model is linear, yet is able to reproduce much of the phenomenology that has required the consideration of non-linear rheology in previous attempts. This is an important conceptual improvement that justifies the consideration of this kind of model. The model is presented in the next Section. Results are contained in Section III, and Section IV contains a summary and the conclusions.

\section{Model}

The model that will be presented here is based on the one that was introduced in \citet{jagla_pre}, and that has been shown to produce realistic sequences of seismic events that quantitatively compare very well with actual seismic sequences. Here I will rephrase the main ingredients of the model, focusing on its application in the context of creep rupture.
I consider a set of $N_0$ linear elastic fibers, characterized by the value of their stiffness $k_i$ ($i=1,...,N_0$). The fibers are considered to be equally loaded, i.e., the same stretching $\varepsilon$ is applied to all of them. Fibers break when a maximum stretching value $\varepsilon^{th}_i$ is reached. Note that for the case of uniform stiffness, namely $k_i\equiv k_0$, the model corresponds to the usual fiber bundle model in the global load shearing (GLS) scheme (also known as local load shearing, or democratic fiber bundle) as discussed extensively by \citet{pradhan_rmp}, for instance. Here I will modify the model presented by \citet{jagla_pre} in the following way. Let us suppose first of all that the fibers have all the same breaking strain $\varepsilon_0$,  but that they are not anchored to the point of coordinate $0$, but instead to some point of coordinate $u_i$. This means that effectively, the stretching that has to be applied to break a particular fiber is $\varepsilon=\varepsilon_0+u_i$. Let us consider now the mechanical energy stored in the system at some given value of $\varepsilon$. This is given by

\begin{equation}
E={\sum_i}' \frac{k_i}{2}(\varepsilon-u_i)^2
\end{equation}
where the prime in the sum means that this is extended only to intact fibers, namely those with $(\varepsilon-u_i)< \varepsilon_0$. The key new point in the system dynamics is to consider that the values of $u_i$ are not constant but they can evolve in time. To work on a safe physical basis, the evolution of $u_i$ will be such that the mechanical energy $E$ tends to be minimized in the process. The simplest dynamics one can assume for such a temporal variation consists in calculating $du_i/dt$ as

\begin{equation}
\frac{du_i}{dt}=-\lambda \frac{\delta E}{\delta u_i}\equiv \lambda f_i= \lambda k_i (\varepsilon-u_i)
\end{equation}
for intact fibers, where I note the force on fiber $i$ as $f_i$. This would correspond to the modeling of each fiber as a linear combination of a harmonic spring (with length $\varepsilon-u_i$) plus a linear dashpot represented by the variable $u_i$. This situation however is not appropriate for the case we are trying to model, since in particular, it implies that for any fixed stretching, eventually the system reaches a configuration of zero stress.\footnote[1]{Note that in the present form this model is similar with the one used by  \citet{baxevanis_2007,baxevanis_2008}, with the difference that they consider that fiber rupture occurs at a maximum value of (in the present naming of variables) $\varepsilon+u_i$. In their case the unrealistic behavior manifests in the fact that the system eventually breaks for any value of applied stress, no matter how small.}
This would correspond to a viscous fluid rather than to the creeping material we are trying to model. Instead of the previous situation in which the temporal variation of $u_i$ is proportional to the force on the fiber $i$ alone, I propose to model creep assuming that the temporal variation of $u_i$ will be proportional to the {\em fluctuation} of the force in the neighborhood of the fiber $i$. In the case all fibers are intact, it is direct to make explicit this prescription in the form of a conserving dynamics, namely

\begin{equation}
\frac{du_i}{dt}=\lambda \nabla^2\frac{\delta E}{\delta u_i}= -[\nabla^2 f]_i=\sum_j(f_i-f_j)
\end{equation}
where the index $j$ labels the four neighbors (on a square lattice) of the site $i$.
Some care is however necessary when there are broken fibers, as they have in some sense come out of the game, and should not contribute to determine the temporal evolution of $u_i$. The way in which this is done is the following. For each intact fiber $i$, the temporal variation of $u_i$ is given by

\begin{equation}
\frac{du_i}{dt}=\lambda \sum_j (f_i-f_j)
\label{dudt}
\end{equation}
where the index $j$ is now understood to run only onto intact neighbor fibers of the site $i$. This is the proposed form of the time evolution of $u_i$ that I will study in detail in this work.
A couple of things are worth to be noted. For a given value of $\varepsilon$, relaxation through Eq. (\ref{dudt}) leads eventually to a situation in which the force onto any connected patch of intact fibers (if any) becomes uniform. This is the {\em relaxed} configuration that the system approaches, asymptotically.
Also, note that if all $k_i$ were equal, starting from a situation in which all $u_i=0$, the previous evolution gives no time dependence of $u_i$. This points to the crucial role played by the fact that in the present model the values of $k_i$ are widely distributed.

I will concentrate mainly in the case in which an external constant force $f$ per fiber is applied to the system. 
When one fiber breaks, its load is immediately distributed onte the intact fibers, and this can produce the immediate failure of some other fibers, generating an avalanche.  Whether there is a stationary condition in which some of the fibers remain intact, or all fibers are eventually broken, depends on the applied load onto the system.  The protocol to obtain the time evolution of the system, in particular, the sequence of fiber breaking is the following. Previously to the application of the force, it is assumed that the system had plenty of time to relax, in such a way that the forces onto each fiber have become equal to zero everywhere. This means that all $u_i=0$. At $t=0$ a nominal force $f$ per fiber is applied. Assuming that initially all fibers are intact, the following sequence of steps is performed (primed sums run only onto intact fibers):

1- The value of $\varepsilon$ is calculated from $\sum '_i(\varepsilon-u_i)k_i=fN_0$

2- Check if any fiber has overpassed the breaking threshold, i.e., if $(\varepsilon-u_i)>\varepsilon_0$. If yes, consider this/these fiber/s as broken, and return to 1.

3- Once no additional broken fibers occur in step 2, increase time in a small time step $dt$, calculate the new $u_i$ values using (\ref{dudt}), and return to step 1.

In this way, given an externally imposed value of the force $f$, the time dependence of the stretching $\varepsilon(t)$ can be followed. $\varepsilon(t)$ will typically have periods of continuous evolution when no fibers breaks, and abrupt jumps occurring when fibers break. I emphasize that as the force carried by any fiber that breaks is distributed onto the intact fibers (through steps 1 and 2), a cascade of events can occur in which a finite number of fibers break exactly at the same time. This is an interesting variant with respect to other viscoelastic fiber bundle model \citep{nechad_prl,kun_pre_2003}, in which the breaking of fibers occurs in a strictly one-by-one sequence. 

A technical comment on the implementation of the previous rules is appropriate here. If we want to follow in full detail the bursts of breaking activity in the system, the time step $dt$ in step 3 above has to be tuned in such a way that it never produces by itself the breaking of more than one fiber at a time. This is easily done as the precise value $dt_0$ necessary to break the next fiber can be calculated due to the linearity of the springs. Then in this approach, the value of $dt$ is chosen dynamically at each passage through step 3. This full detail process however, is very time consuming if we are only interested in the overall evolution of $\varepsilon$ with time. In this case, a fixed value of $dt$ is used, and in this time interval a few different fibers can break independently, however, the overall evolution of $\varepsilon(t)$ is very well reproduced.
Now I will present detailed results on the numerical implementation of this model.

\section{Results}

\subsection{Results for average quantities}

I already indicated that a broad distribution of stiffness $k_i$ is necessary for the model to be sensible. The precise form of this distribution is however not crucial for the observed properties. I will present results for a uniform distribution of $k$ between 0 and $k_{max}\equiv 1$. Additional calculations using a Weilbull distribution have shown no appreciable qualitative differences.

\begin{figure}[h]
\centerline{\includegraphics[width=.5\textwidth]{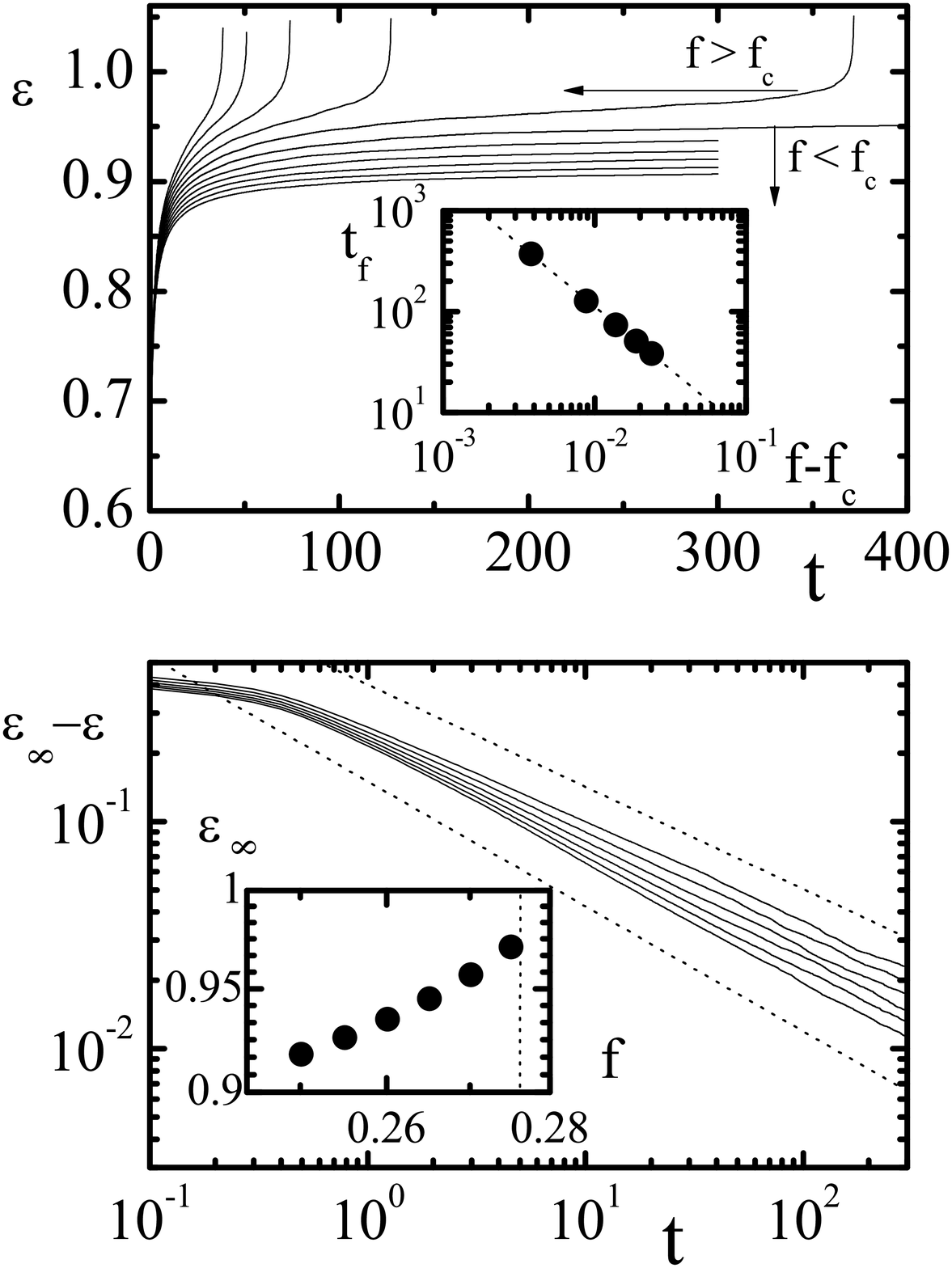}}
\caption{(a) Strain as a function of time at different values of the applied load
in a single realization on a system of size 1000 $\times$ 1000. In the inset, the time to failure $t_f$ for $f>f_c$ is plotted as a function of load, showing a power law divergence approaching $f_c=0.2763$. Points are the numerical results, dotted line has a slope of -1.25.
(b) The approach of $\varepsilon$ to its asymptotic value $\varepsilon_{\infty}$ for the case $f<f_c$ is shown, and is seen to follow a power law behavior (dotted lines have slopes -0.45 and -0.55, for reference). The best determined asymptotic values of the strain as a function of the applied force are plotted in the inset, where the critical force $f_c$ is marked as a vertical dotted line.
}
\label{f1}
\end{figure}

Depending on the value of the external load per fiber $f$, the configuration of the system at a very large time corresponds to one of two different possibilities. If $f$ is lower than some critical value $f_c$ only a fraction of fibers break, and the value of $\varepsilon$ remains bounded as $t\to \infty$, whereas if $f>f_c$, all fibers break in a finite time.  

\begin{figure}[tbp]
\centerline{\includegraphics[width=.5\textwidth]{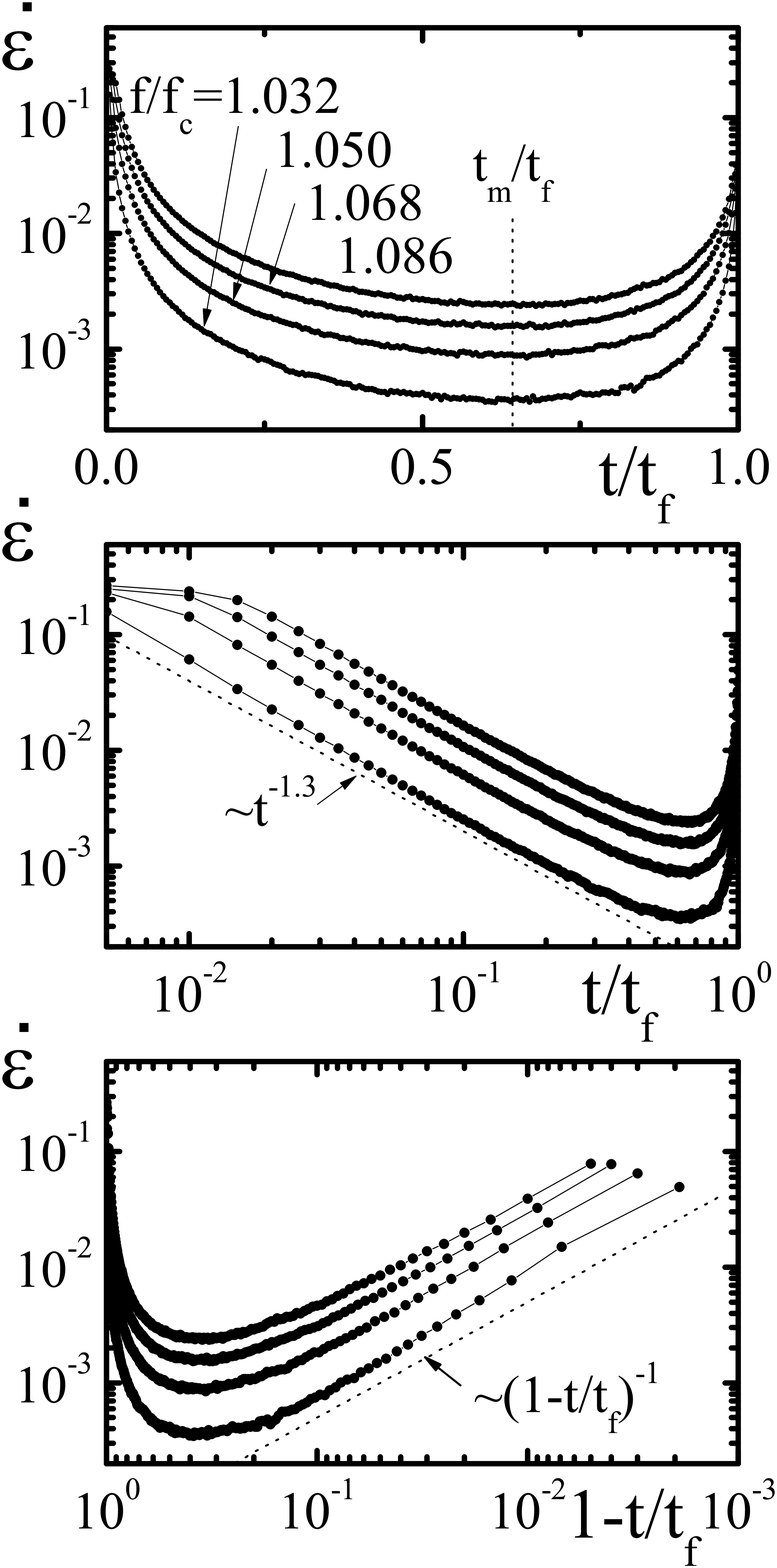}}
\caption{Strain rates $\dot \varepsilon \equiv d\varepsilon/dt$ as a function of time normalized to time-to-failure in the run away regime ($f>f_c$). The data shown correspond to an average over 120 realizations in systems of size 1000 $\times$ 1000. In first panel time scale is linear, and the time of  minimum strain rate has been emphasized. In the second panel time scale is logarithmic, to display the power law decay in the primary creep regime (note however the larger-than-one exponent compared to usual Andrade creep). In the last panel a logarithmic time scale approaching $t_f$ is used, to emphasize the power law increase of strain rate as approaching failure.
}
\label{f2}
\end{figure}

In Fig. \ref{f1}(a) we see the evolution of the stretching $\varepsilon$ as a function of time, for a single realization in a system of 1000 $\times$ 1000 fibers. The qualitative difference between the two regimes $f>f_c$ and $f<f_c$, is clearly observable. For the case $f>f_c$, the time to failure in the system diverges as $(f-f_c)^{-\alpha}$, with $\alpha\simeq 1.25 \pm 0.01$, as can be seen in the inset. For the present realization, $f_c$ turned out to be $f_c=0.2763$. This value shows rather large fluctuations in different realizations. 
In panel (b) of Fig. \ref{f1}, I show the approaching of $\varepsilon$ to the asymptotic values, where a power law dependence is clear. The asymptotic values have been adjusted to obtain the best power law fitting, and they are shown as a function of $f$ in the inset. It is seen that they follow a smooth behavior as $f_c$ is approached, without obvious signs of any singular behavior. 
The power law approach of  $\varepsilon$ to the asymptotic values is compatible with the Andrade creep law. Note however that the exponent $\simeq 0.5$ corresponds to a dependence of the strain {\em rate} $\dot \varepsilon$ of the form $\dot \varepsilon \simeq 1/t^{1.5}$, whereas the usual exponent for strain rate in the Andrade regime is about 2/3. Note however that the usual Andrade regime for $f-f_c$ has eventually to cross over to a different dependence that guarantees a finite strain at infinite time, (otherwise the lower-than-one exponent would produce a divergent strain). Here the larger-than-one exponent obtained gives a finite strain at infinite time.

To make a better comparison with the results presented by \citet{nechad_prl}, in Fig. \ref{f2} I present results for $f>f_c$, showing the values of strain rate in the system $\dot \varepsilon\equiv d\varepsilon/dt$, as a function of $t/t_f$, under different scaling of the temporal axis. In the first panel the temporal axis is linear, and rescaled with the time to failure $t_f$. Here we see that $\dot \varepsilon$ displays qualitatively the three mayor creeps stage before rupture: a primary regime of decreasing creep rate, a second regime of almost constant creep, and a tertiary regime of accelerated creep towards failure.

From this figure we see that the time at which the creep rate is minimum is between at 0.6-0.7 of $t_f$, independently of $f$, a result that was observed to occur systematically in real samples, and that was not captured by the models available up to now. The logarithmic temporal scale in panel (b) of Fig. \ref{f2}, shows that the first stage of creep in the model corresponds indeed to an Andrade regime of potential decay of creep rate with time. The exponent observed for this decay is seen to be about 1.3,
which is larger than the usually observed exponent that is typically near 2/3. The logarithmic plot with respect to $(t_f-t)$ in panel (c) contains a very interesting outcome of the model: strain rate shows a tendency to diverge at $t_f$ as $(t_f-t)^{-1}$, which is in fact the result that is observed in a variety of experimental cases. It has to be reminded here, that fiber bundle models provide typically a divergence of strain rate as $(t_f-t)^{-1/2}$, whereas the model by \citet{nechad_prl,nechad_jmps} obtains a systematic $(t_f-t)^{-1}$ dependence relying on a specific non-linear choice for the rheology of the fibers. 

An interesting additional point on the ``divergence" of $\dot \varepsilon$ at $t_f$ has to be noticed. Although the plot in Fig. \ref{f2}(c) shows an increase of $\dot \varepsilon$ by about two orders of magnitude compatible with a $(t_f-t)^{-1}$ dependence, this behavior cannot be extrapolated to $t\to t_f$. In fact, if this was the case, then the value of $\varepsilon$ itself would diverge at $t_f$ but this is nonsense: there are no fibers that can withstand an arbitrarily large deformation. The situation is that $\dot \varepsilon$ increases as $(t_f-t)^{-1}$ (and $\varepsilon$ as $1/\log(t_f-t)$) up to some time slightly before $t_f$, at which a macroscopic event breaks all remaining fibers (which is a finite fraction of the total system) all of the sudden. We will confirm this scenario in the next section where I concentrate on the study of bursts of activity in the system.

\subsection{Burst sizes distribution}

The results in Figs. \ref{f1} and \ref{f2} focus on the average behavior of the system, namely, the evolution of the stretching $\varepsilon$ or its time derivative $\dot \varepsilon$ on time, and are indications that the model is able to capture in a quite realistic way the time evolution of strain that is observed in materials experiencing creep rupture. Now we will see that fluctuation in these quantities, associated in particular to the existence of discrete breaking events, reveal also very interesting behaviors that can be compared for instance with results of experiments on acoustic emission.

\begin{figure}[h]
\centerline{\includegraphics[width=.5\textwidth]{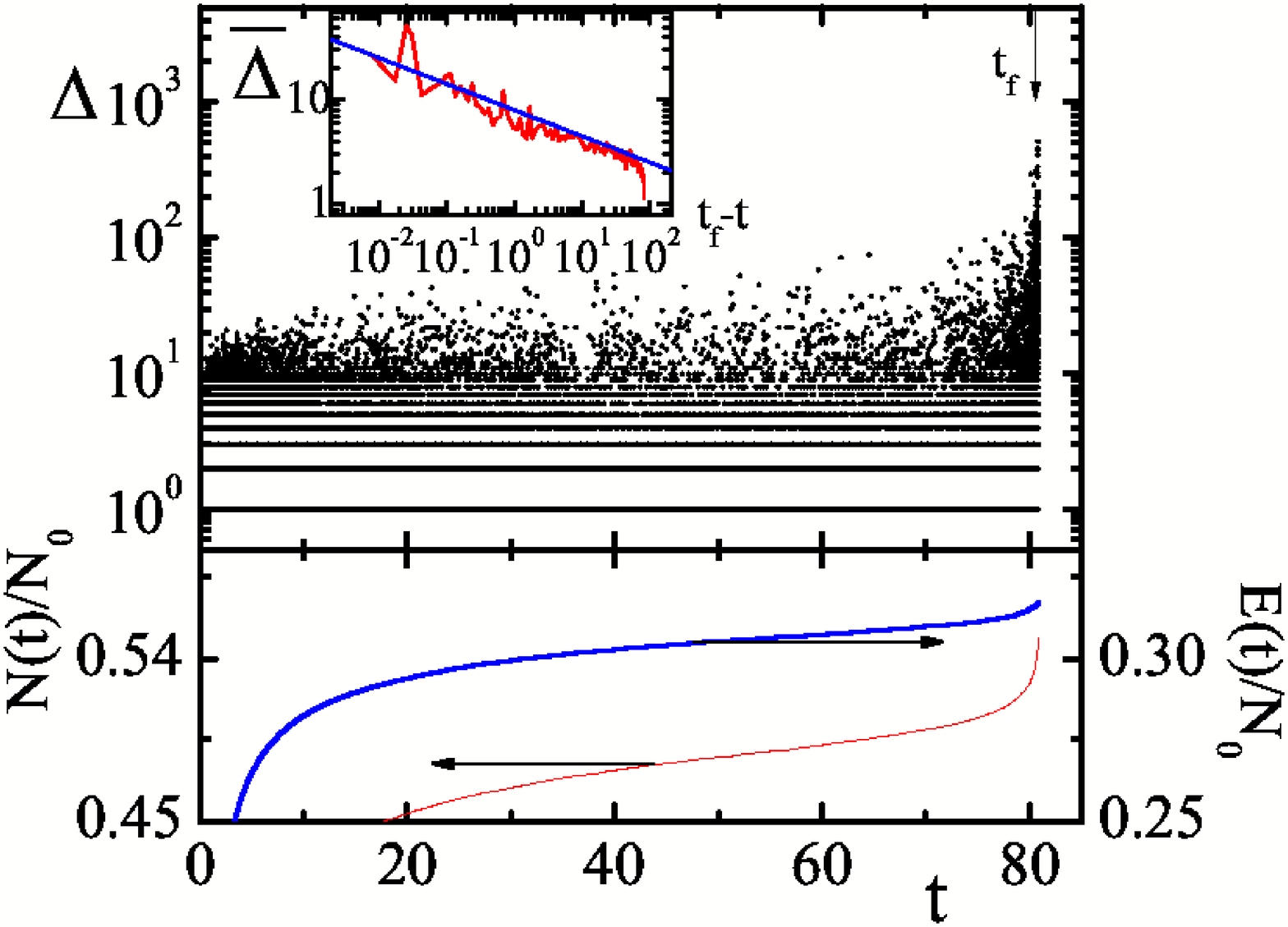}}
\caption{Main upper panel: A sequence of all breaking events in a system of size 1000 $\times$ 1000, for a load slightly larger than critical ($f/f_c=1.05$). The average burst size $\overline \Delta$ as a function of time is indicated in the inset, in log scale (the power law of the straight line has an exponent 0.25). The cumulative number of events $E(t)$, and the cumulative number of broken fibers $N(t)$ as a function of time is shown in the panel below. 
}
\label{f3}
\end{figure}

Some previous work have focused on the size distribution of bursts. However, in some other models the burst do not occur exactly as a function of time under a constant force as in our case. In the much studied static model considered in particular by \citet{pradhan_rmp}, it is assumed that the external force increases quasi-statically in time, then the bursts occur because $f$ is increasing, and not directly because of the existence of creep phenomena. Anyway this is a case that has been studied in much detail, and is the one with which we should compare, also because there are interesting analytical predictions for the size distribution  of bursts in this case. In fact, it is known that if bursts are counted during the whole load process, from $f=0$ to complete failure, the number of bursts of size $\Delta$ follows a power law of the form $\Delta \sim \Delta ^{-\xi}$, with $\xi=5/2$. If however, the size of the bursts is collected only in a small time interval before breaking occurs, the exponent predicted is $\xi=3/2$. 
Some other viscoelastic models, as the one by \citet{kun_pre_2003} produce a one-by-one breaking of fibers, and in that case a non-trivial size distribution of events cannot be defined. The model of \citet{baxevanis_2007} seems to be more realistic in this respect, and the $\xi=5/2$ and $\xi=3/2$ power laws for the total and close-to-rupture avalanches have been obtained in this case also.

The model we are studying here gives a non trivial time sequence of bursts for any fixed vale of $f$.
In Fig. \ref{f3} we see a typical time sequence of all breaking events in the system, for a value of $f$ slightly above the critical value $f_c$. 
The cumulative number of broken fibers $N(t)$ is qualitatively very similar to the evolution of $\varepsilon(t)$ discussed in the previous section. This number is seen to approach  a fraction of about 0.55 of $N_0$ as $t$ approaches $t_f$. At this point, a macroscopic breaking event occurs, in which the rest of the system breaks. Note that the fraction of fibers that breaks in this single event (about $0.45 N_0$) is rather independent of the value of $N_0$ itself, and on the value of $f$ (assumed larger than $f_c$).
The existence of this macroscopic breaking event is reasonable in the present model, where the breaking of a fiber implies the immediate load increase on all the other fibers. In a more realistic situation, particularly as the time of global failure is approaching, even very small delays in the process of redistribution of stresses will have an important influence on the dynamics of failure. I expect that this consideration results in a continuous evolution of $N(t)$ towards $N_0$ at $t_f$ in that case.
The results for the the average size $\overline \Delta $ of the bursts as a function of time (inset in Fig. \ref{f3}) shows that $\overline \Delta $
increases as a power law as $t_f$ is approached, namely $\overline \Delta (t) \sim 1/(t-t_f)^z$, with $z\simeq 0.25$. 

\begin{figure}[h]
\centerline{\includegraphics[width=.5\textwidth]{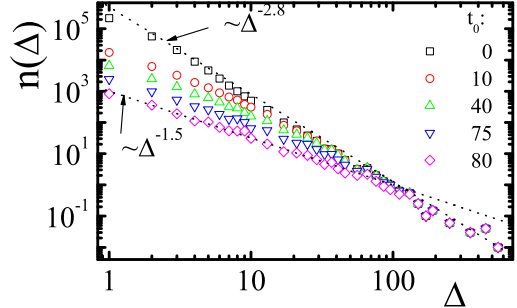}}
\caption{Distribution of bursts for the sequence in figure \ref{f3}. Results are presented for different time intervals, namely $[t_0, t_f]$ for different values of $t_0$ ($t_f\simeq 80.82$ for this particular realization). The distribution is seen to pass from a power law with a decaying exponent $\sim 2.8$ for the full distribution (i.e., $t_0=0$), to a $1.5$ power law when $t_0 \to t_f$.
}
\label{f4}
\end{figure}

The distribution of bursts corresponding to the sequence in Fig. \ref{f3} is shown in Fig. \ref{f4} for different time interval $[t_0, t_f]$ with $0<t_0<t_f$. The complete burst distribution ($t_0=0$) shows a consistent power law with a decaying exponent around 2.8. As $t_0$ increases, a crossover and a coexistence of two different power laws is observed. For $t_0\to t_f$ the distribution tends to be described by a single power law with an exponent close to 1.5. These results are qualitatively similar (except for a slightly larger global exponent) to those obtained in the models studied by \citet{pradhan_rmp} and \citet{baxevanis_2008}. 

\begin{figure}[h]
\centerline{\includegraphics[width=.5\textwidth]{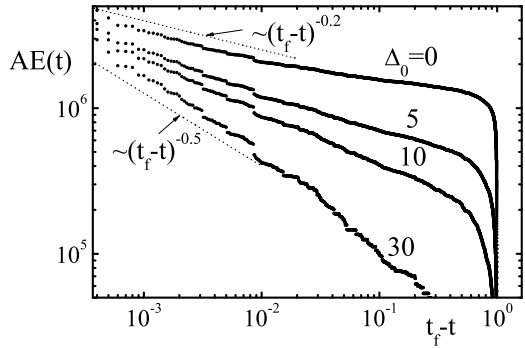}}
\caption{Cumulative energy released as would be detected in an acoustic emission experiment corresponding to the sequence of events shown in Fig. \ref{f3}. Different curves correspond to different detectability threshold $\Delta_0$, as indicated. A tendency to increase as a power law as $t_f$ is approached is observed, however, the exponent is dependent on the threshold used. Two lines with exponents 0.2 and 0.5 are drawn for comparison.
}
\label{fae}
\end{figure}

Results for bursts sizes can be cast is a slightly different form in order to compare with available results on acoustic emission. In particular, one of the magnitudes that has been experimentally recorded is the cumulative energy released as a function of time, on approaching failure. It has been obtained \citep{garcimartin,ciliberto} that this magnitude shows a power law increase on approaching failure with a dependence of the form $\sim (t_f-t)^\gamma$, with reported values for $\gamma$ close to 0.25. The experimental detection and processing of an acoustic signal involves some non-trivial steps that make the link with other physical properties of the system not a straightforward procedure. However, for the present model we can reasonably consider that a burst involving the breaking of $\Delta$ fibers produces an acoustic emission signal of energy $\sim \Delta^2$, as the energy is experimentally defined as the time integral of the square of the amplitude detected \citep{ae,nechad_jmps}. Also, we should consider that events involving less that some minimum value $\Delta_0$ of broken fibers are not detected because lack of sensitivity. In Fig. \ref{fae} I show the result for the accumulated energy released, calculated by integrating in time the values of $\Delta^2$, for events with $\Delta$ larger than a threshold $\Delta_0$, corresponding to the sequence of events shown in Fig. \ref{f3}. The results show the right tendency to diverge as $t\to t_f$. The definition of a well defined power law $\sim (t_f-f)^{-\gamma}$ is however strongly dependent on the threshold $\Delta_0$ that is used. Values of $\gamma$ between 0.2 and 0.5 can be obtained by changing $\Delta_0$.

\subsection{Interevent times}

\begin{figure}[h]
\centerline{\includegraphics[width=.5\textwidth]{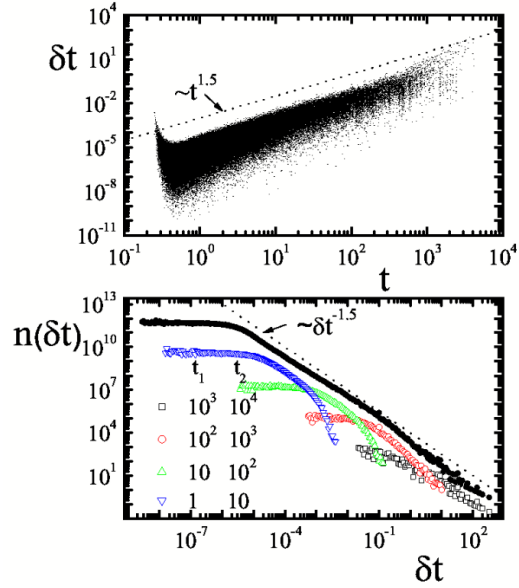}}
\caption{(a) A sequence of interevent times $\delta t$ in a single realization in a 1000 $\times$ 1000 system, for a load slightly below the critical value. The average interevent time increases approximately as $t^{1.5}$ (dotted line). (b) Full symbols: distribution of interevent times in  the complete time interval $[0,\infty]$. Open symbols: distribution of interevent times for different  time intervals $t_1$-$t_2$, as indicated. 
}
\label{f5}
\end{figure}

Another quantity that is expected to provide information on microscopic processes in the system, and that is experimentally accessible in acoustic emission measurements is the distribution of inter event times (IETs), namely, the time intervals $\delta t$ between successive breaking events in the system. For systems in which bursts occur in an uncorrelated time sequence, an exponential (Poissonian) distribution of IETs is expected. Deviations from the exponential distribution might be taken as signs of non-trivial correlations in the activity of the system. However, this interpretation needs to be used with care, at least.
In fact, in Fig. \ref{f5}(a) we see the full sequence of and IETs for a case of an applied load slightly below the critical value. The activity in the system decays in time towards a stationary state, and this is seen in the systematic increase of IET with time. The distribution if IETs for this case is shown in part (b) of Fig. \ref{f5}. It is seen there that IETs show a power law distribution over more than seven orders of magnitude. The origin of this
power law is however trivially related to the fact that the average IET as a function of time is itself a power law increasing function of time.
By calculating the distributions of IETs for partial time intervals (shown in Fig. \ref{f5}(b)) makes clear that the power law appears only as a result of the integration of different time windows.

\begin{figure}[h]
\centerline{\includegraphics[width=.5\textwidth]{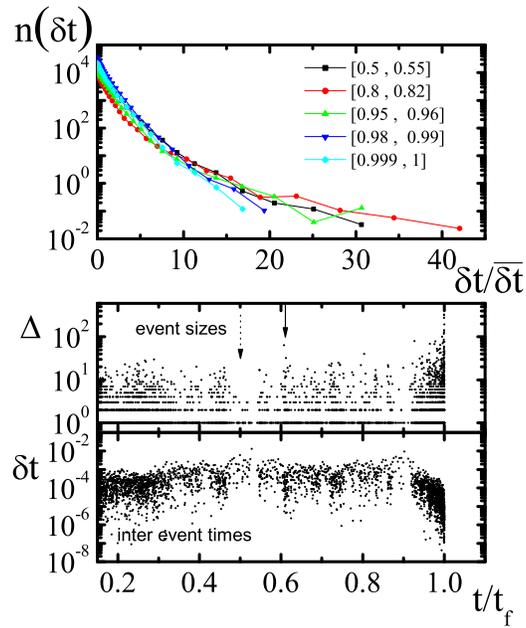}}
\caption{
(a) Distribution of IETs within small time windows $[t_1/t_f, t_2/t_f]$ as indicated, in systems of 400 $\times$ 400, for $f$ slightly abobe $f_c$ (the result of an average over 120 realizations is shown). 
The data do not follow an exponential distribution, as systematic deviations of a straight line plot are observed. This indicates temporal clustering of the events. 
(a) A particular sequence of events and IETs where the temporal clustering is observed. A temporal region of enhanced (depressed) activity is highlighted by the continuous (dotted) arrow.}
\label{f6}
\end{figure}

It is thus evident that if non-trivial correlations between events exist, their signatures have to be searched within small time windows, in which the process can be considered stationary. This fact seems not to have been taken into account properly in some previous studies of IETs distributions.
In Fig. \ref{f6} I show results in a system of 400 $\times$ 400 for a force slightly above the critical value. A smaller system size is used in this case as it makes easier to detect correlations between events. 
In order to have a reasonable statistics, I run different realizations, and the results have been combined by scaling the occurrence time of the events by the corresponding failure time $t_f$ of that particular realization. 
The results for the distribution of IETs calculated for small time windows (Fig. \ref{f6}(a)), show a clear departure from the Poissonian behavior. The curvature of the data indicates an increased probability of events to occur at small or large time intervals, whereas probability for intermediate time intervals
is depressed. This indicates a tendency of the data to cluster in time, which is also pretty clear in the individual realization shown in the lower part of the figure. Temporal clustering of events in the present model is reminiscent of temporal clustering of earthquakes, in the form of aftershocks. In the present model, temporal clustering of events is a direct consequence of the relaxation mechanism present in the model, as it was qualitatively explained in the version of the model devised to study seismic phenomena \citep{jagla_pre,jagla_jgr}. This kind of clustering should be detectable in acoustic emission experiment.

\section{Conclusions}

In the present paper a model to describe the creep rupture of materials has been presented. The model is in the class of viscoelastic fiber bundle models, and the viscoelastic relaxation mechanism considered involves a tendency to locally uniformize the stresses within the system. The model is a slight modification of one that has been proved useful to explain many characteristics of the sequences of events (earthquakes) in a seismic context.
The only non-linear ingredient of the model is the possibility that fibers break when a maximum stretching is overpassed, in particular, I do not rely upon any non-linear rheology of the elements of the system.
Using this model, I have obtained a description of the creep rupture process that in a few respects is an improvement with respect to the results in previous modeling. In particular, a regime of primary (Andrade) creep has been obtained, in which the strain rate has a power law dependence. The exponent of this dependence, is however larger than one, contrary to the usually observed exponent around 2/3. A region of accelerated creep preceding rupture was obtain, in which the strain rate increases as $\sim (t_f-t)^{-1}$, where $t_f$ is the time of failure. The time of minimum strain rate is systematically observed to occur at about 0.6 of the time of failure, coincident with experiments.
Bursts sizes distributions around rupture with an exponent close to 3/2, and for the whole temporal sequence with an exponent slightly above 5/2 where obtained, much similar to results with previous modeling. The accumulated acoustic emission energy in the model was shown to increase as a power law as the time of failure is approached. The observed exponent is however dependent of the cut off used as a threshold for detecting events, and varies between 0.2 and 0.5. A power law distribution of interevent times (IETs) is trivially originated in a temporal variation of the average IET. Analysis of IETs in time windows of constant average activity reveal that the activity is correlated to some extent, displaying temporal clustering of the events. This was indicated to be reminiscent of the aftershocks that are produced with this model when used in the context of seismic phenomena.


\section{Acknowledgments}

This research was financially supported by Consejo Nacional de Investigaciones Cient\'{\i}ficas y T\'ecnicas (CONICET), Argentina. Partial support from
grant PICT 32859/2005 (ANPCyT, Argentina) is also acknowledged.


\end{document}